\documentclass[onecollarge,natbib]{svjour2}
\bibpunct{[}{]}{;}{n}{}{,} 
\smartqed  
\usepackage{graphicx}
\usepackage{epstopdf}
\usepackage{braket}
\journalname{Archive of Applied Mechanics}
\begin{document}

\title{Study of Twist-2 GTMDs in scalar-diquark model
}


\author{Satvir Kaur         \and
        Harleen Dahiya 
}


\institute{Satvir Kaur \at
              Dr. B. R. Ambedkar National Institute of Technology, Jalandhar, Punjab-144011, India. \\
              \email{satvirkaur578@gmail.com}           
           \and
           Harleen Dahiya \at
              Dr. B. R. Ambedkar National Institute of Technology, Jalandhar, Punjab-144011, India.
}

\date{Received: date / Accepted: date}

\maketitle

\begin{abstract}
We investigate the Generalized Transverse Momentum-dependent Distributions (GTMDs) describing the parton structure of proton using the light-front scalar-diquark model. In particular, we study the Wigner distributions for unpolarized quark in unpolarized, longitudinally-polarized and transversely-polarized proton. We also investigate the twist-2 GTMDs for unpolarized quark $(\Gamma=\gamma^+)$ in scalar-diquark model.
\end{abstract}

\section{Introduction}
\label{intro}
One of the most interesting aspect of quantum chromodynamics (QCD) is to get the knowledge about the relationship between partons (quarks and gluons) and hadrons. There are different parton distributions present to describe the relationship. Parton distribution function $f(x)$ describes the probability of finding a parton carrying a longitudinal momentum fraction $(x)$. To get the information about the distribution of parton in transverse direction of the motion of hadron, generalized parton distributions (GPDs) were introduced \cite{meibner}. GPDs are accessible in hard exclusive processes like deeply virtual Compton scattering (DVCS) or meson production. Transverse momentum-dependent parton distributions (TMDs) give explanation of parton distributions in momentum plane which can be measured in certain semi-inclusive reactions like semi-inclusive deep inelastic scattering (SIDIS) or the Drell-Yan (DY) process \cite{meibner}. GPDs and TMDs represent the three-dimensional picture of hadron.
\par The quark and gluon Wigner distributions were introduced by Ji \cite{ji}, to better understand the hadron structure. Wigner distributions are joint position and momentum distributions which is the quantum analog to the classical phase-space distributions. These distributions have been investigated in light-front dressed quark model, light-cone chiral soliton model, AdS/QCD inspired quark-diquark model, light-cone spectator model etc. \cite{tmaji, jmore}. Wigner distributions can be studied as a fourier transformation of generalized transverse momentum-dependent parton distributions (GTMDs). GTMDs are entitled as \textit{mother distributions} of GPDs and TMDs. One can derive GPDs and TMDs by integrating GTMDs upon quark transverse momentum $(\textbf{p}_\perp)$ and by applying the forward limit in the hadron momentum ($\Delta=0$) respectively. There are 16 complex-valued GTMDs at leading twist, which contain the rich information about the hadron structure. 
\section{Quark Wigner distributions}
One can obtain the Wigner distributions $\rho^{[\Gamma]}({\bf b_\perp},{\bf p_\perp},x;S)$ by taking the two-dimensional Fourier transform from $\bf{\Delta}_{\perp}$ to the impact-parameter space coordinates $\textbf{b}_{\perp}$ of quark-quark correlator (Wigner operator) $W^{[\Gamma]}({\bf \Delta_\perp}, {\bf p_\perp},x;S)$ \cite{jmore, gpcf}. We have
\begin{eqnarray}
\rho^{[\Gamma]}({\bf b_\perp},{\bf p_\perp},x;S)&=& \int \frac{d^2 \bf \Delta_\perp}{(2 \pi)^2} e^{-i {\bf{\Delta}_\perp} \cdot {\bf b_\perp}} W^{[\Gamma]}({\bf \Delta_\perp}, {\bf p_\perp},x;S),
\label{wigner}
\end{eqnarray}
where $W^{[\Gamma]}({\bf \Delta_\perp}, {\bf p_\perp},x;S)$ in the proton state $\Ket{P;S}$ at fixed light-cone time $z^{+}=0$ is defined as
\begin{eqnarray}
W^{[\Gamma]}({\bf \Delta_\perp}, {\bf p_\perp},x;S)&=&\frac{1}{2} \int \frac{dz^- d^2 z_\perp}{(2 \pi)^3} e^{i p \cdot z} \Bra{P'';S} \bar{\psi}(-z/2) \Gamma \mathcal{W}_{[-\frac{z}{2},\frac{z}{2}]} \psi(z/2)\Ket{P';S}\Bigm\vert_{z^{+}=0},
\end{eqnarray}
with $P'$ and $P''$ be the initial and final momenta of proton, $S$ be the proton spin and $\Gamma$ indicates the Dirac matrix ($\gamma^{+}$, $\gamma^{+}\gamma_{5}$, $i\sigma^{j+}\gamma_{5}$, where $j$ = 1 or 2). The gauge link Wilson line, $\mathcal{W}_{[-\frac{z}{2},\frac{z}{2}]}$ ensures the $SU(3)$ color gauge invariance of the Wigner operator.
\par We use light-front scalar-diquark model to evaluate Wigner distributions and GTMDs, which is assumed to be the bound state of a quark and scalar-diquark having spin-0. 
The LCWFs for spin up state of proton are \cite{Baccheta}
\begin{eqnarray}
\psi_{+}^{+}(x, \textbf{p}_{\perp})&=&(m+x M)\frac{\phi}{x},\\
\psi_{-}^{+}(x, \textbf{p}_{\perp})&=&(p_{x}+ip_{y})\frac{\phi}{x}.
\end{eqnarray}
The LCWFs for spin down state of proton are
\begin{eqnarray}
\psi_{+}^{-}(x, \textbf{p}_{\perp})&=&-[\psi_{-}^{+}((x, \textbf{p}_{\perp})]^{*},\\
\psi_{-}^{-}(x, \textbf{p}_{\perp})&=&\psi_{+}^{+}(x, \textbf{p}_{\perp}),
\end{eqnarray}
where $
\phi(x, \textbf{p}_{\perp})=-\frac{g_{s}}{\sqrt{1-x}}\frac{x(1-x)}{\textbf{p}_{\perp}^{2}+ (xM_{s}^{2}+(1-x)m^{2}-x(1-x)M^{2})}.$
\par There are 16 independent twist-2 quark Wigner distributions followed by the various polarization configurations. Here, we evaluate 3 Wigner distributions of unpolarized quark explicitly from Eqn (\ref{wigner}).\\
Wigner distribution of unpolarized quark in an unpolarized proton \cite{nk, lorce} is given as
\begin{eqnarray}
\rho_{UU}({\bf b_\perp},{\bf p_\perp},x)&=& \frac{1}{2} \Big[\rho^{[\gamma^+]}({\bf b_\perp},{\bf p_\perp},x;+\hat{S}_z) + \rho^{[\gamma^+]}({\bf b_\perp},{\bf p_\perp},x;-\hat{S}_z)\Big],\nonumber\\
 &=& \frac{1}{16\pi^{3}\left( 2\pi\right)^{2}}\int d\Delta_{x} d\Delta_{y} \int dx \ \cos\left( \Delta_{x} b_{x}+\Delta_{y} b_{y}\right) \nonumber\\
&\times &\frac{1}{x^{2}} \left[ \left( \textbf{p}_{\perp}^{2}-\frac{\left( 1-x\right) ^{2}}{4}\bf{\Delta}_{\perp}^{2}\right) +\left( m+xM\right) ^{2}\right]
 \phi^\dagger \left(\bf p_{\perp}^{''}\right) \phi\left(\bf p_{\perp}^{'}\right). 
\label{uu}
\end{eqnarray}
Wigner distribution of an unpolarized quark in longitudinally-polarized proton is given by
\begin{eqnarray}
\rho_{LU}({\bf b_\perp},{\bf p_\perp},x)&=&\frac{1}{2}\Big[\rho^{[\gamma^+]} ({\bf b_\perp},{\bf p_\perp},x;+\hat{S}_z) - \rho^{[\gamma^+]} ({\bf b_\perp},{\bf p_\perp},x;-\hat{S}_z)\Big],\nonumber\\
&=& \frac{1}{16\pi^{3}\left(2\pi\right)^{2}}\int d\Delta_{x} d\Delta_{y} \int dx \ \sin\left( \Delta_{x} b_{x}+\Delta_{y} b_{y}\right) \nonumber\\
&\times &\frac{\left( 1-x\right) }{x^{2}}\left( \Delta_{x} p_{y}-\Delta_{y} p_{x}\right) \ \phi^\dagger \left(\bf p_{\perp}^{''}\right) \phi\left(\bf p_{\perp}^{'}\right).
\label{lu}
\end{eqnarray}
Wigner distribution of unpolarized quark in transversely-polarized proton is given by
\begin{eqnarray}
\rho^i_{TU}({\bf b_\perp},{\bf p_\perp},x) &=& \frac{1}{2} \Big[\rho^{[\gamma^{+} ]} ({\bf b_\perp},{\bf p_\perp},x;+\hat{S}_i) - \rho^{[\gamma^{+}]} ({\bf b_\perp},{\bf p_\perp},x;-\hat{S}_i)\Big],\nonumber\\
\rho^1_{TU}\left( \textbf{b}_{\perp}, \textbf{p}_{\perp}, x\right) &=& -\frac{2}{16 \pi^{3}\left(2\pi\right)^{2}}\int d\Delta_{x} d\Delta_{y} \int dx \ \cos \left( \Delta_{x} b_{x}+\Delta_{y} b_{y}\right)\nonumber\\
&\times & \frac{\left( 1-x\right)}{x^{2}} \Delta_{x} \left( m+xM\right)\phi^\dagger \left( \bf p_{\perp}^{''}\right) \phi\left(\bf p_{\perp}^{'}\right),
\label{tu}
\end{eqnarray}
where
$\textbf{p}^{'}_{\perp}=\textbf{p}_{\perp}-\left( 1-x\right)\frac{\bf{\Delta}_\perp}{2}$ and $
\textbf{p}^{''}_{\perp}=\textbf{p}_{\perp}+\left( 1-x\right)\frac{\bf{\Delta}_\perp}{2}$ are initial and final momenta of quark.
\section{Generalized transverse momentum-dependent parton distributions (GTMDs)}
The GTMDs can be extracted from the different Wigner distributions. There are total 16 complex-valued GTMDs. Here, we studied GTMDs with unpolarized quark i.e. when $\Gamma=\gamma^+$, which relates Wigner quark-quark correlator as \cite{gpcf}
\begin{eqnarray}
 W_{\lambda \lambda'}^{[\gamma^+]}
 &=& \frac{1}{2M} \, \bar{u}(p', \lambda') \, \bigg[
      F_{1,1}
      + \frac{i\sigma^{i+} k_\perp^i}{P^+} \, F_{1,2}
      + \frac{i\sigma^{i+} \Delta_\perp^i}{P^+} \, F_{1,3} \nonumber\\*
 & &  + \frac{i\sigma^{ij} k_\perp^i \Delta_\perp^j}{M^2} \, F_{1,4}
     \bigg] \, u(p, \lambda)
     \,. \label{e:gtmd_1}
     \end{eqnarray}
     GTMDs can be explicitly evaluated as
     \begin{eqnarray}
     F_{1,1}( x, \bf{\Delta}_{\perp}, \textbf{p}_{\perp})&=&\frac{1}{16\pi^{3}}\left[ \left( \textbf{p}_{\perp}^{2}-\frac{\left( 1-x\right) ^{2}}{4}\bf{\Delta}_{\perp}^{2}\right) +\left( m+xM\right) ^{2}\right]
      \phi^\dagger \left(\bf p_{\perp}^{''}\right) \phi\left(\bf p_{\perp}^{'}\right),\\
F_{1,2}(x, \bf{\Delta}_{\perp}, \textbf{p}_{\perp})&=&0,\\
F_{1,3}(x, \bf{\Delta}_{\perp}, \textbf{p}_{\perp})&=&\frac{F_{1,1}}{2}-\frac{2}{16\pi^{3}}M \left(m+x M\right)\frac{\left( 1-x\right)}{x^{2}} \phi^\dagger \left(\bf p_{\perp}^{''}\right) \phi\left(\bf p_{\perp}^{'}\right),\\
\label{f13}
F_{1,4}(x, \bf{\Delta}_{\perp}, \textbf{p}_{\perp})&=&\frac{1}{16\pi^{3}}M^{2}\frac{\left( 1-x\right)}{x^{2}}\phi^\dagger \left(\bf p_{\perp}^{''}\right) \phi\left(\bf p_{\perp}^{'}\right).
\end{eqnarray}
\section{Results and discussion}
\begin{figure}
\centering
\begin{minipage}[c]{0.98\textwidth}
(a)\includegraphics[width=.3\textwidth]{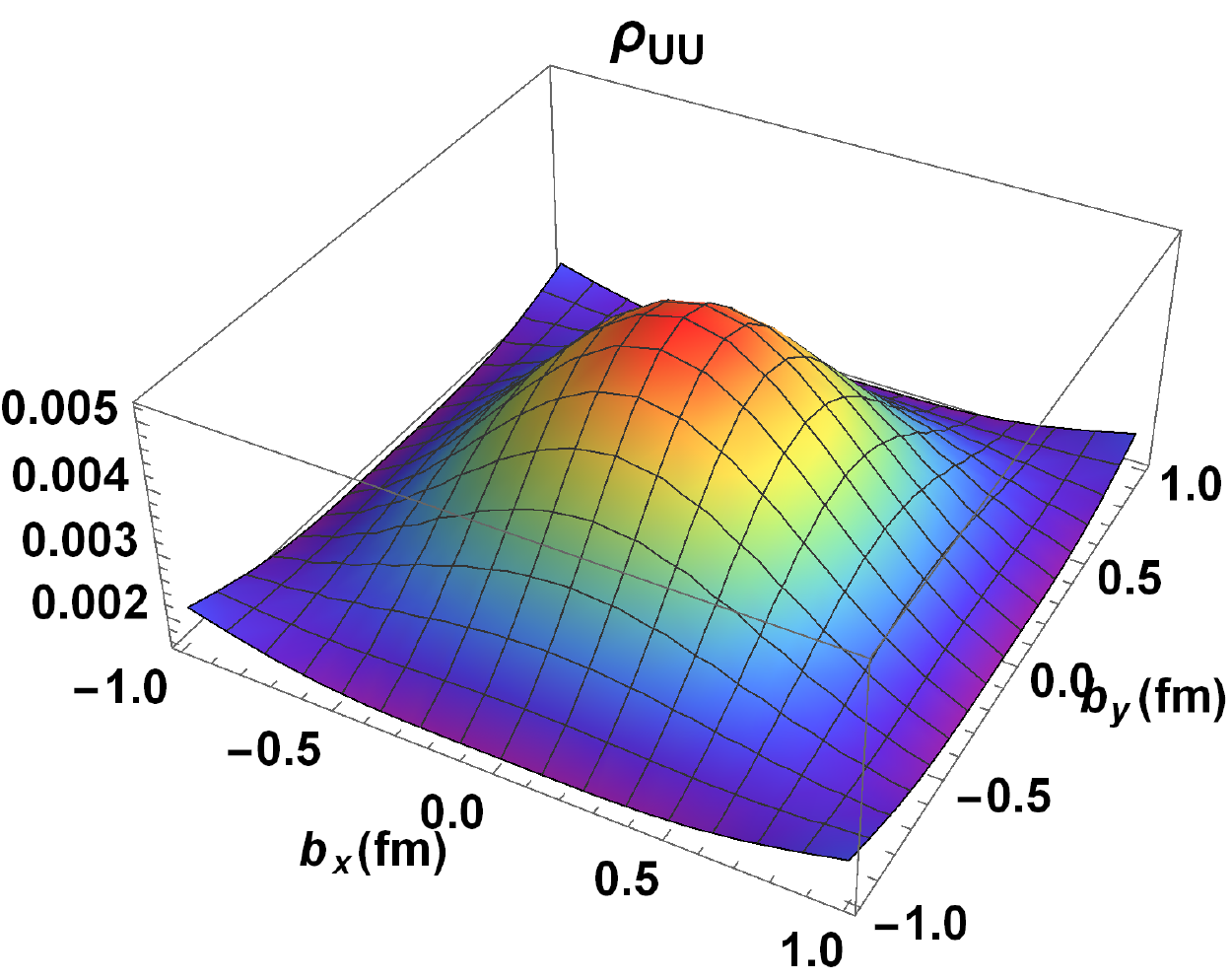}\hfill
(b)\includegraphics[width=.3\textwidth]{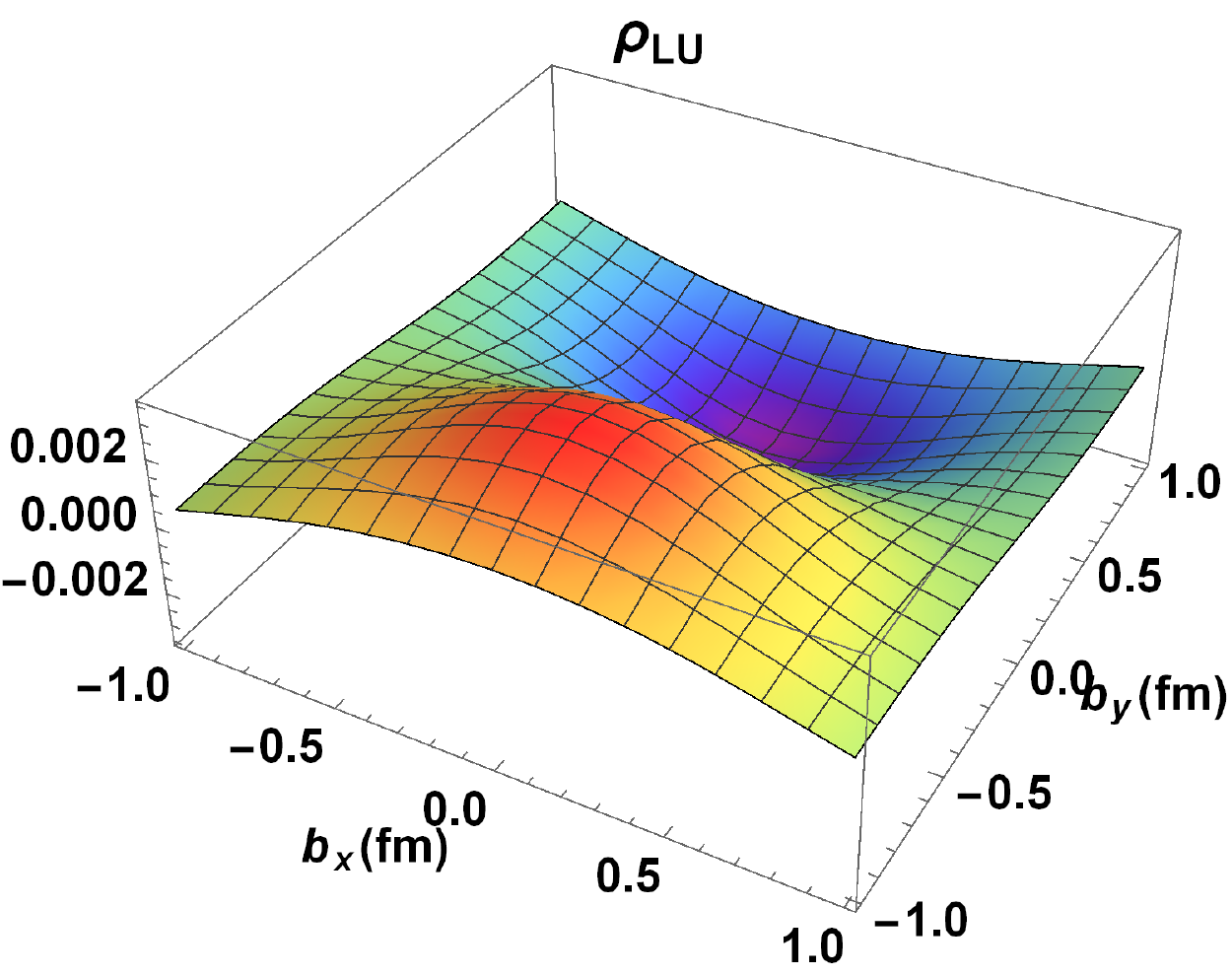}\hfill
(c)\includegraphics[width=.3\textwidth]{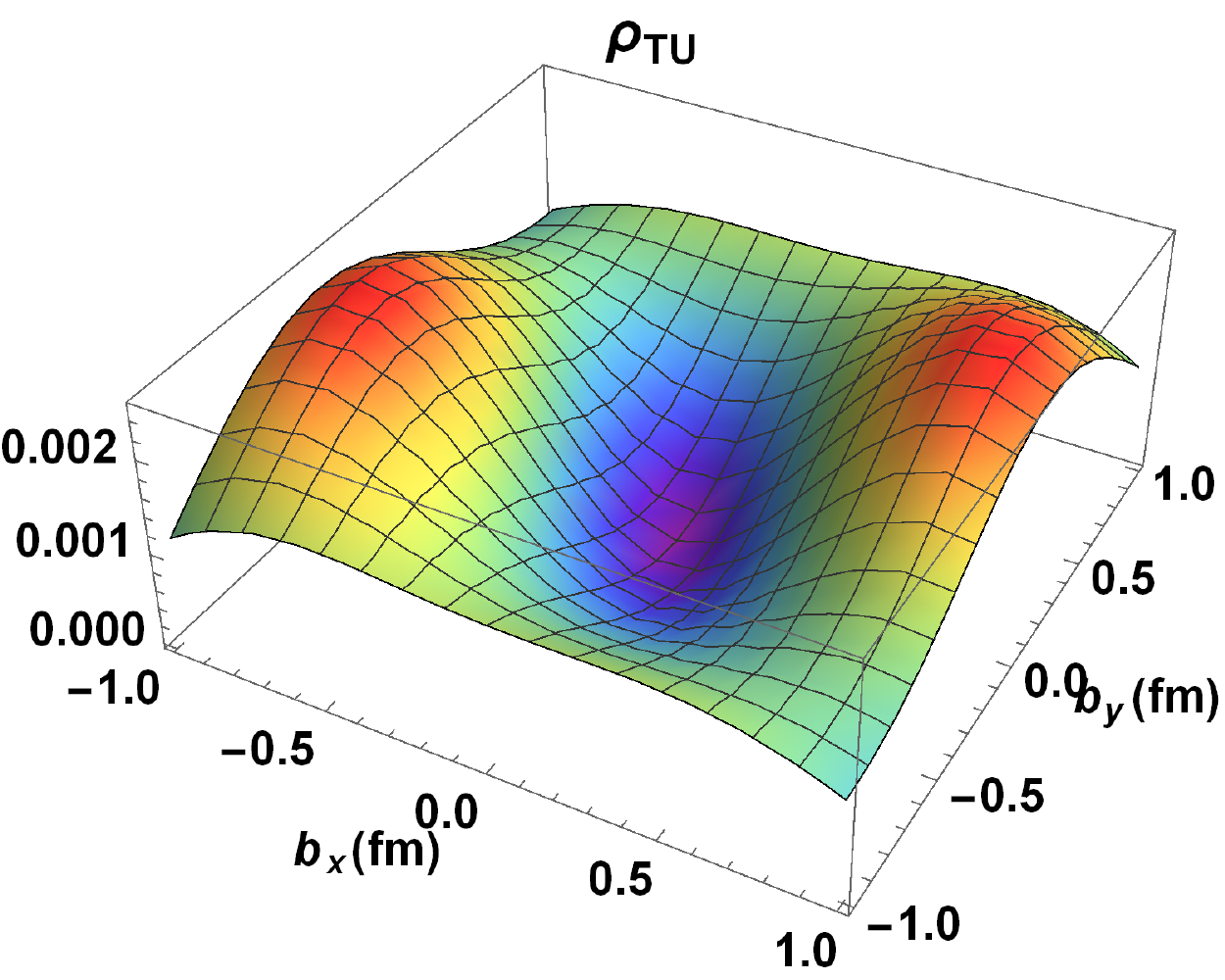}
\end{minipage}
\begin{minipage}[c]{0.98\textwidth}
(d)\includegraphics[width=.3\textwidth]{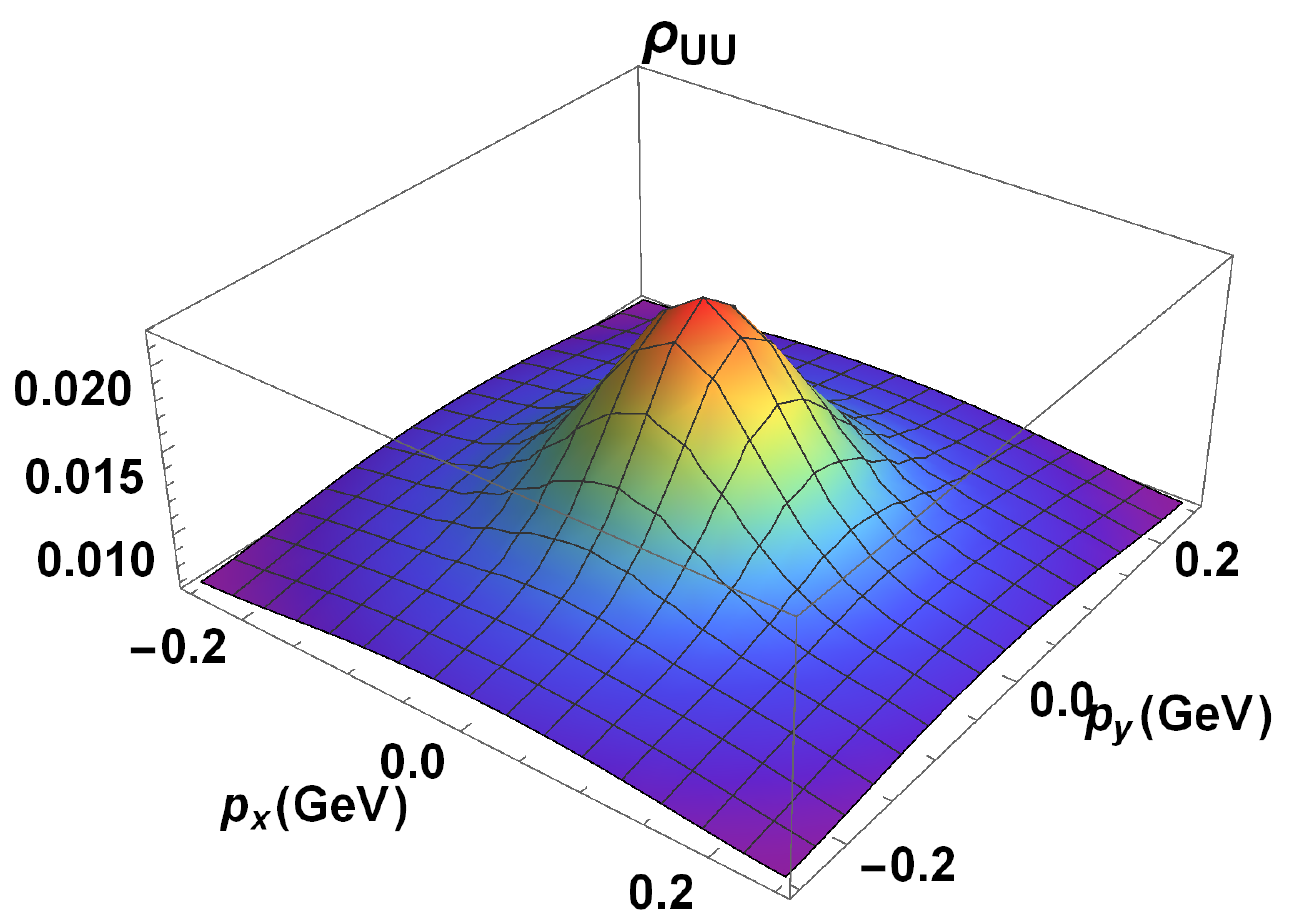}\hfill
(e)\includegraphics[width=.3\textwidth]{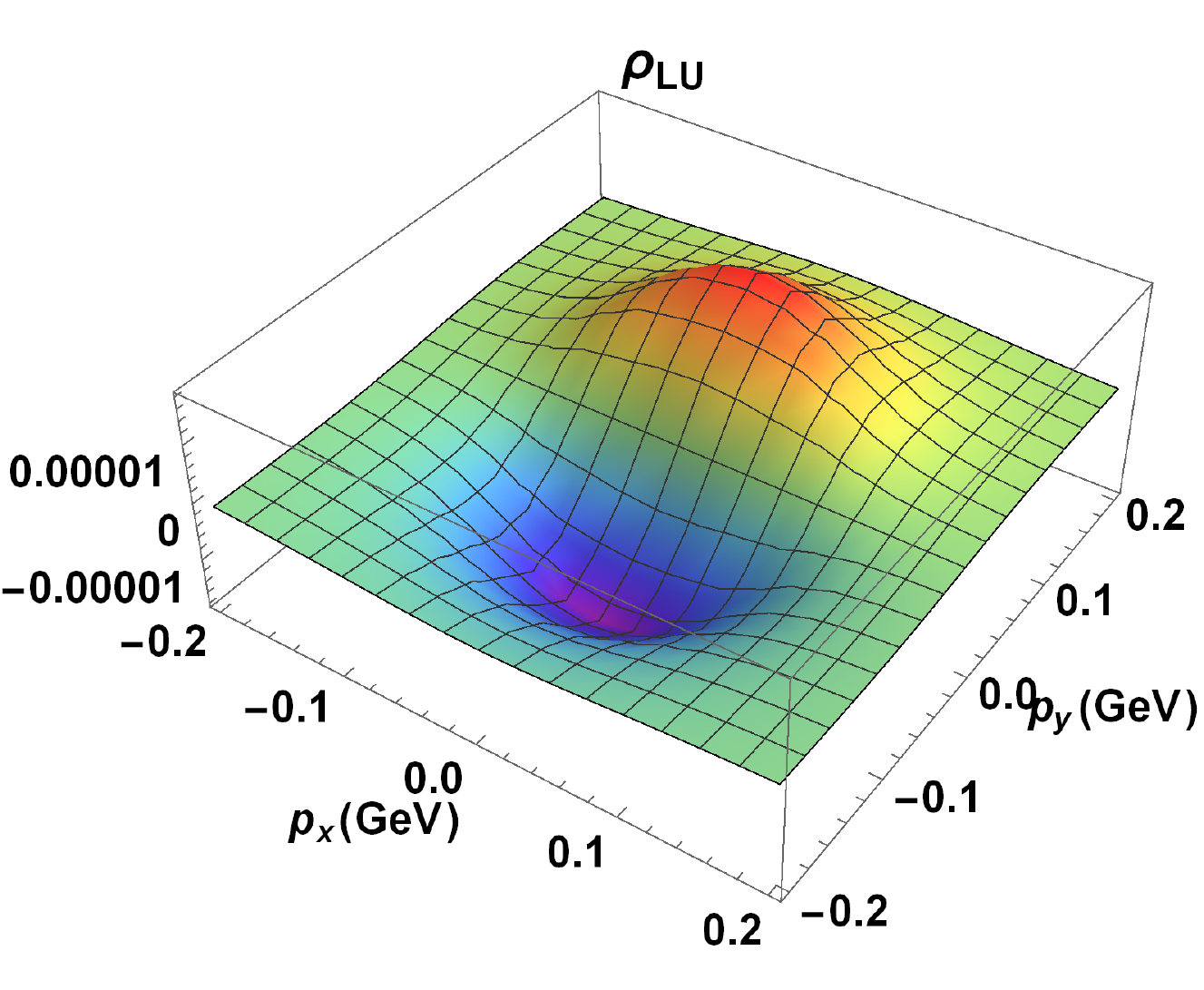}\hfill
(f)\includegraphics[width=.3\textwidth]{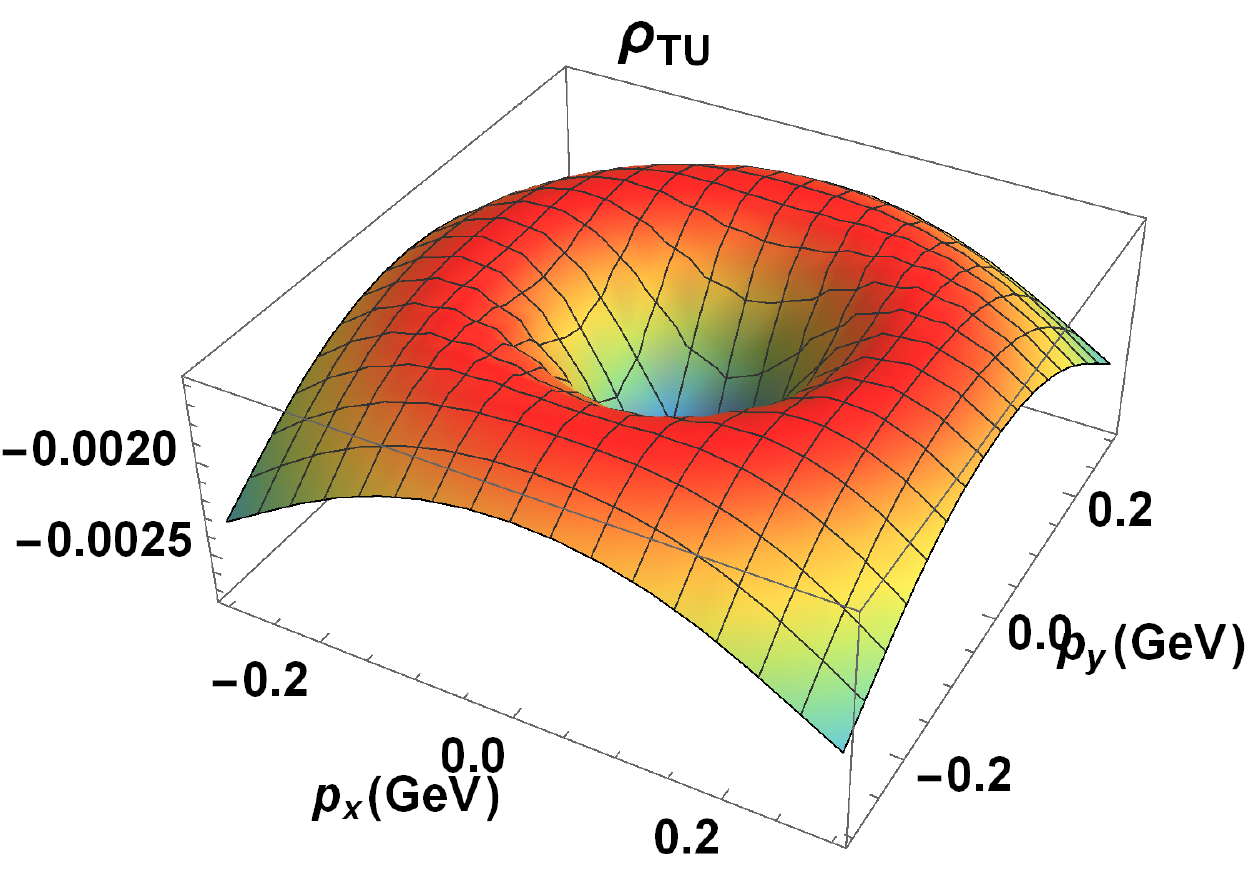}
\end{minipage}
\caption{Quark Wigner distributions $\rho_{UU}({\bf b_\perp},{\bf p_\perp},x)$, $\rho_{LU}({\bf b_\perp},{\bf p_\perp},x)$ and $\rho^i_{TU}({\bf b_\perp},{\bf p_\perp},x)$ in impact-parameter space and momentum space.} 
\label{bbpp}
\end{figure}

In Figs. \ref{bbpp}(a) and \ref{bbpp}(d), we plot unpolarized Wigner distributions in impact parameter space at fixed value of quark momentum $\textbf{p}_{\perp}$ along the direction $\widehat{x}$ i.e. at $p_{x}=0.3$ $GeV$ and in momentum space at fixed value of impact-parameter $\textbf{b}_{\perp}$ along the direction $\widehat{x}$ i.e. at $b_{x}=0.4$ $fm$ respectively. In this model, the distributions are circularly symmetric in both the spaces. The peaks of distribution $\rho_{UU}$ in impact-parameter space and momentum space are at the center and slowly decrease towards edges. The distribution is more concentrated at the center in momentum space. At GPD and TMD limit i.e. by integrating the distributions over $\textbf{b}_{\perp}$, the probabilistic distributions named GPDs and TMDs can be obtained. In this case, at TMD limit, $\rho_{UU}$ corresponds to the T-odd TMD $f_{1}$. 
\par The Wigner distribution of unpolarized quark in longitudinally-polarized proton $\rho_{LU}$ in impact-parameter space and in momentum space are shown in Figs. \ref{bbpp}(b) and \ref{bbpp}(e) respectively. We observe that longitudinally-unpolarized Wigner distribution at fixed $p_{x}=0.3$ $GeV$ and $b_{x}=0.4$ $fm$ indicates the dipolar behaviour in impact-parameter space and momentum space respectively. But the dipolar polarity is opposite in both spaces. No TMD and GPD can be derived from longitudinal-unpolarized Wigner distribution viz. $\rho_{LU}$. This distribution is an important implication for orbital angular momentum.
\par The Wigner distribution of unpolarized quark in transversely-polarized proton i.e. $\rho^i_{TU}$ in transverse impact-parameter space $\textbf{b}_\perp$ and in momentum space $\textbf{p}_{\perp}$ are shown in Figs. \ref{bbpp}(c) and \ref{bbpp}(f). We take the polarization of proton along $\hat{x}$ direction. In impact-parameter space, we see an asymmetry which is not in any specific direction while in momentum space, it shows a circularly symmetric behaviour with the peak in negative direction. From Eq. (\ref{tu}), we observe the strong correlation between the proton spin direction and the parallel transverse co-ordinate. At GPD limit, $\rho_{TU}$ relates to both $E$ and $H$ GPDs and at TMD limit, this distribution obtains Sivers function $f_{1T}^{\perp}$.

\par
\begin{figure}
\centering
\begin{minipage}[c]{1.0\textwidth}
(a)\includegraphics[width=.3\textwidth]{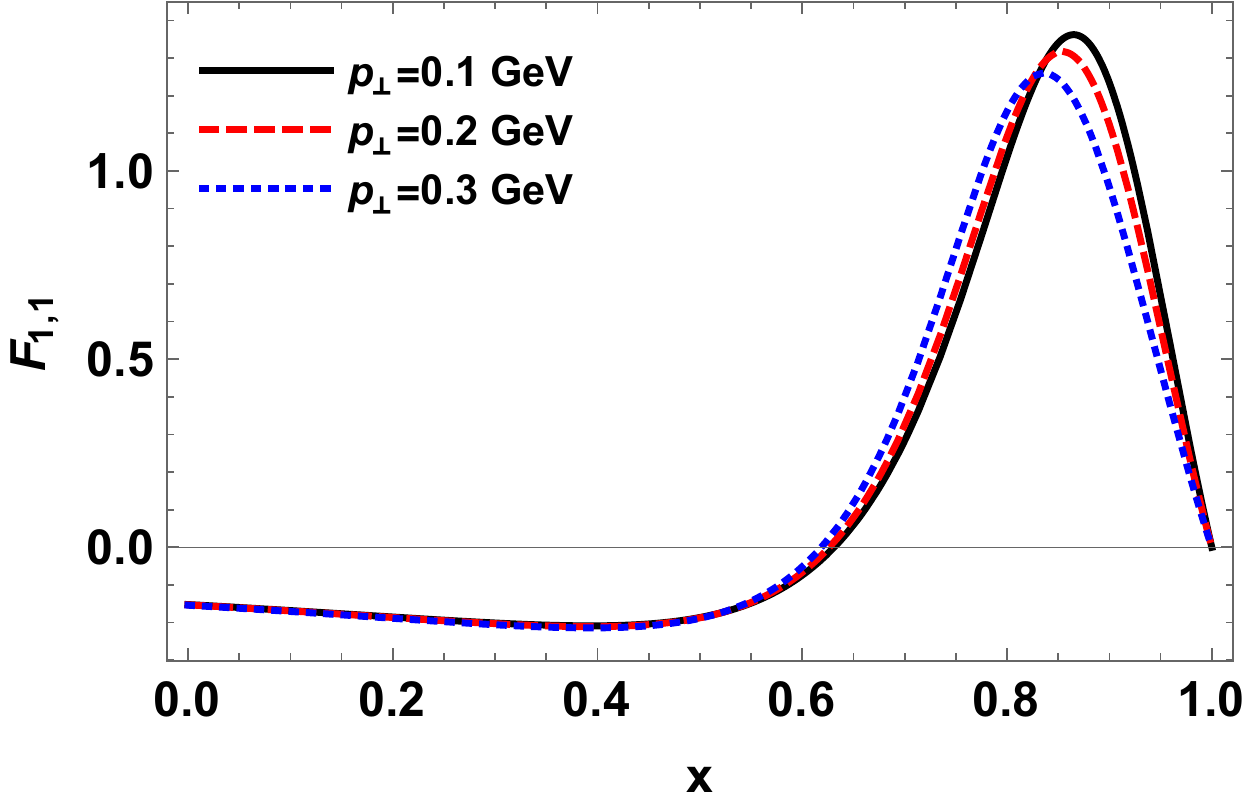}\hfill
(b)\includegraphics[width=.3\textwidth]{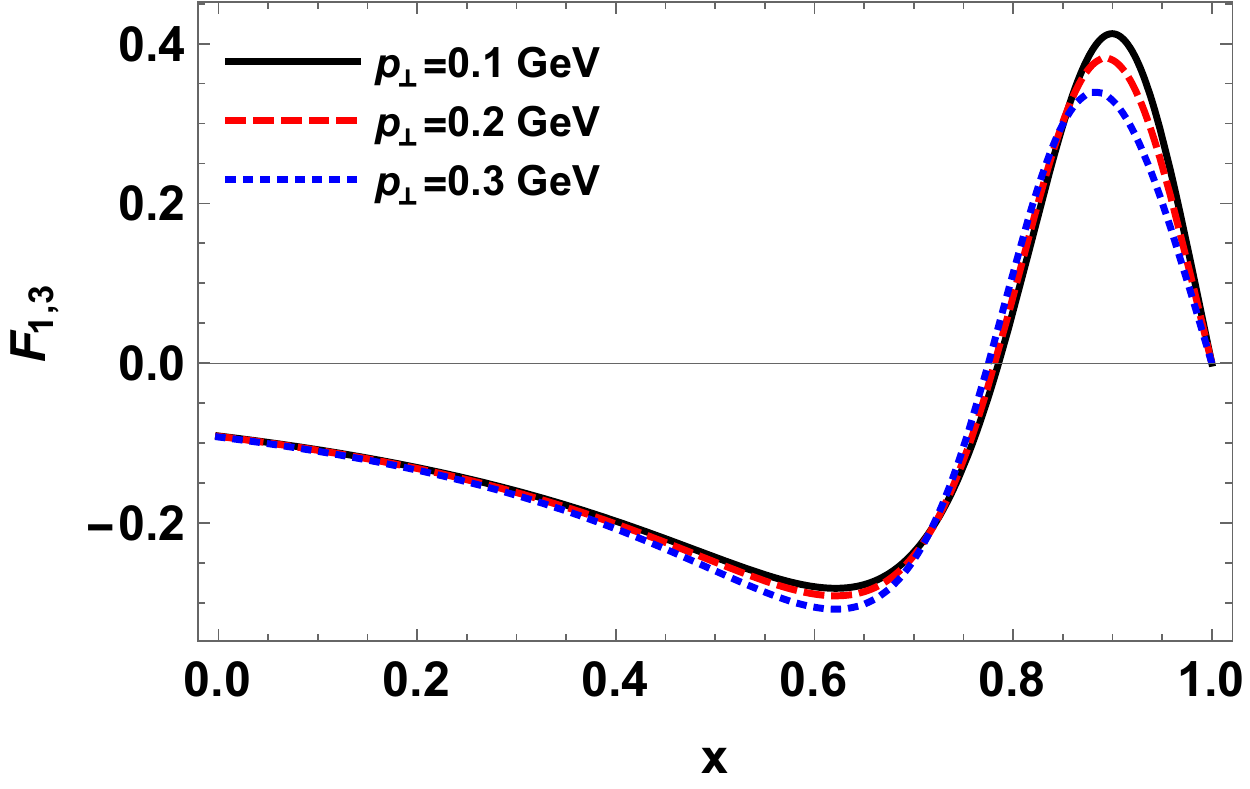}\hfill
(c)\includegraphics[width=.3\textwidth]{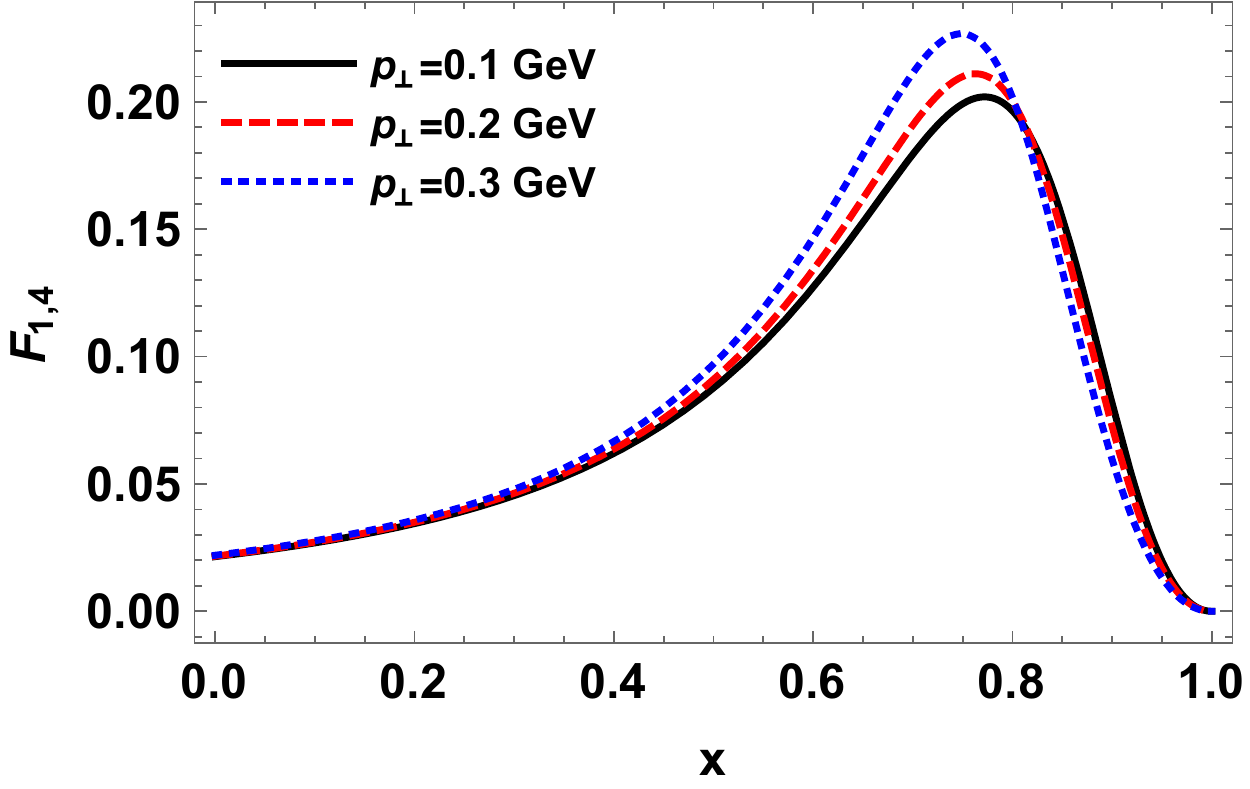}
\end{minipage}
\begin{minipage}[c]{1.0\textwidth}
(d)\includegraphics[width=.3\textwidth]{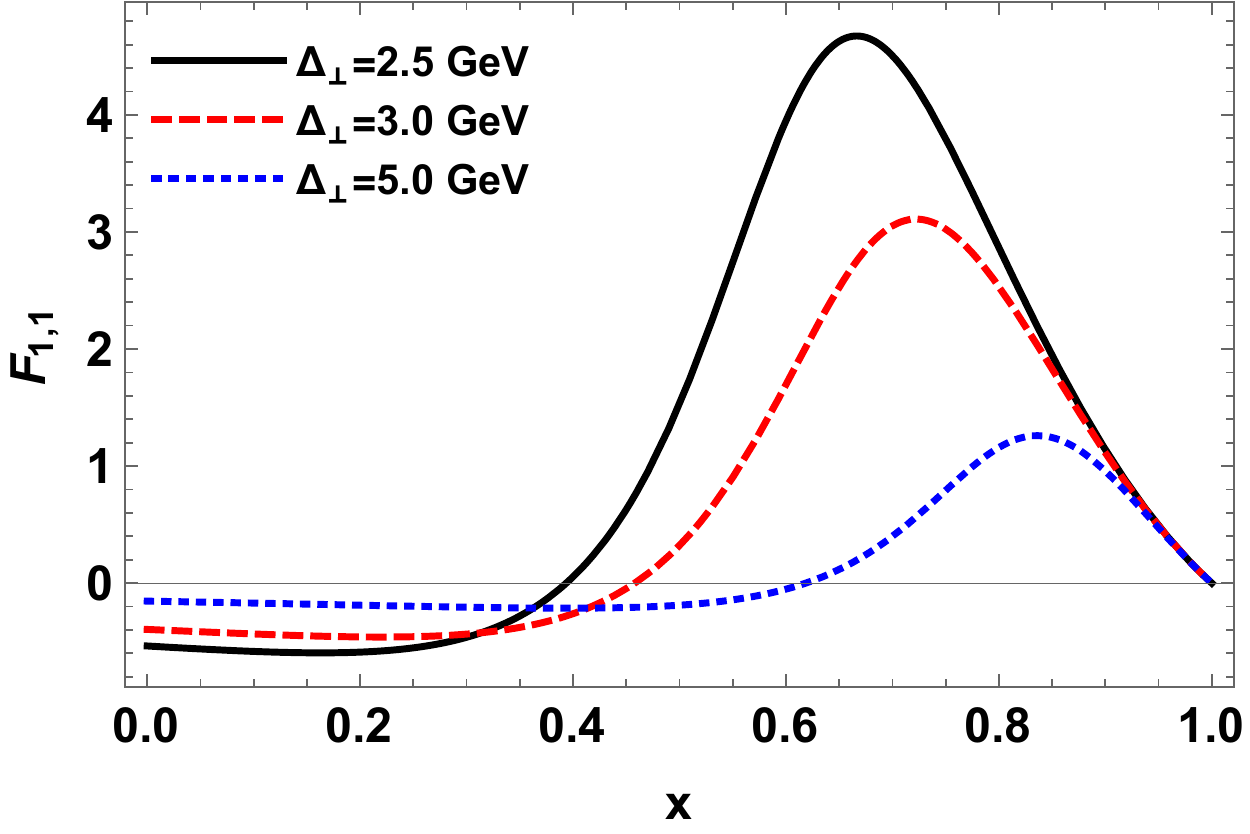}\hfill
(e)\includegraphics[width=.3\textwidth]{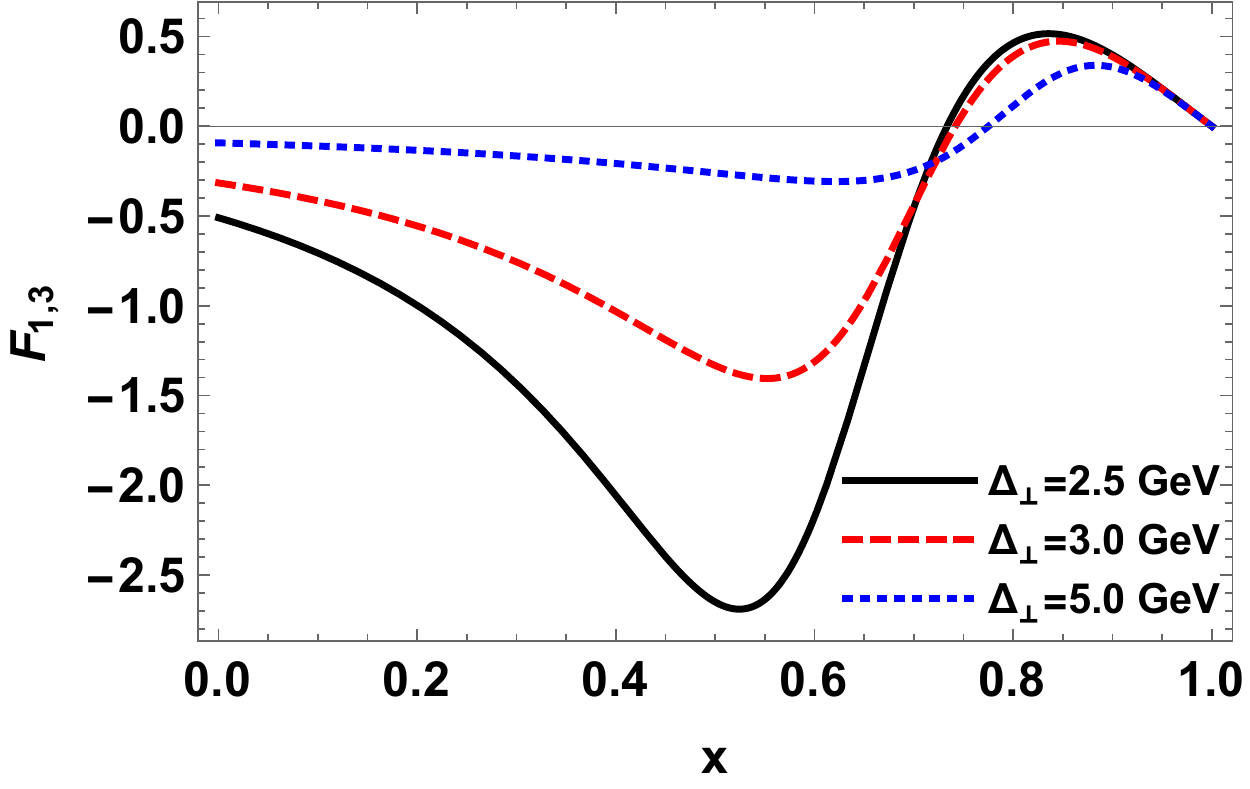}\hfill
(f)\includegraphics[width=.3\textwidth]{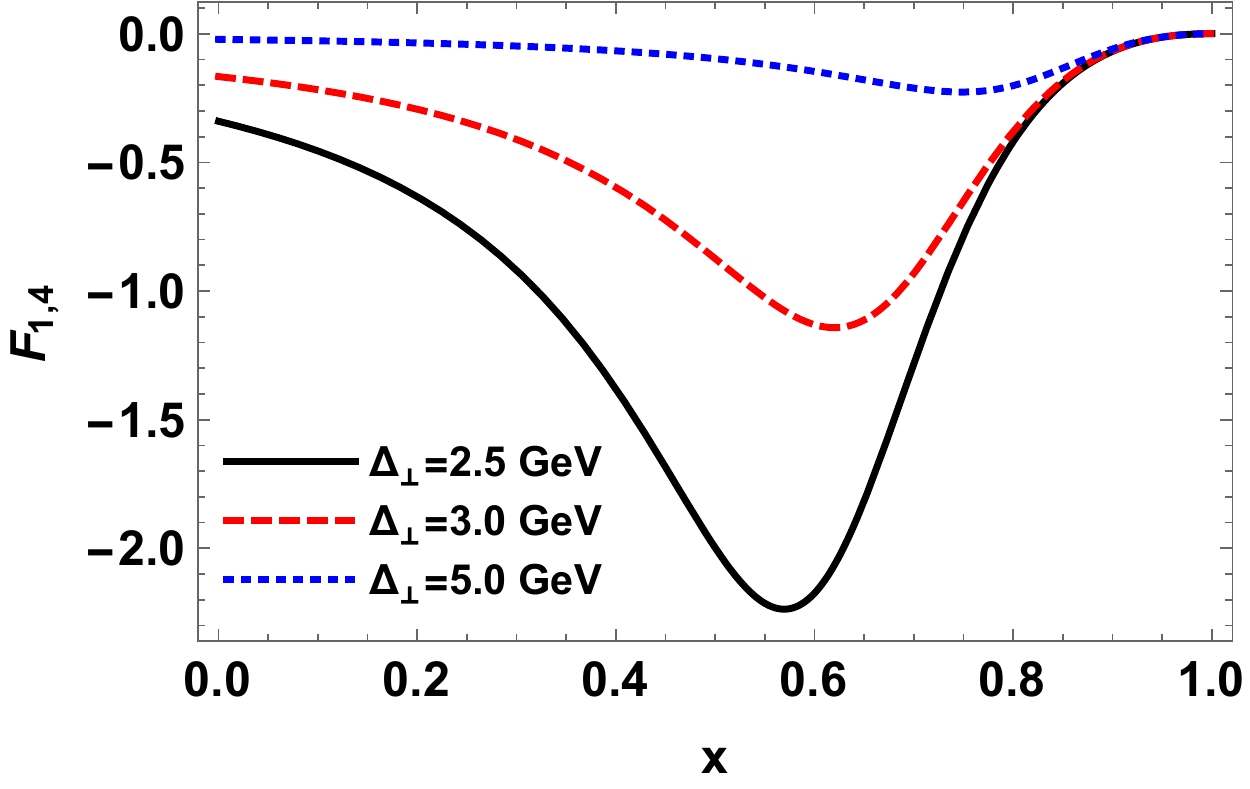}
\end{minipage}
\caption{GTMDs $F_{1,1}(x, \bf{\Delta}_{\perp}, \textbf{p}_{\perp})$, $F_{1,3}(x, \bf{\Delta}_{\perp}, \textbf{p}_{\perp})$ and $F_{1,4}(x, \bf{\Delta}_{\perp}, \textbf{p}_{\perp})$ at fixed $\bf \Delta_{\perp} $= $5$ $GeV$ with the different values of $\textbf{p}_{\perp}$ (upper panel) and at fixed $\textbf{p}_{\perp}=0.3$ $GeV$ with the different values of $\bf \Delta_{\perp} $ (lower panel).}
\label{gtmds}
\end{figure}

 In Figs. \ref{gtmds}(a), \ref{gtmds}(b) and \ref{gtmds}(c), we plot GTMDs $F_{1,1}$, $F_{1,3}$ and $F_{1,4}$ at fixed value of $\bf{\Delta}_{\perp}$ = $5$ $GeV$ with different values of quark momentum $\textbf{p}_\perp$ and in Figs. \ref{gtmds}(d), \ref{gtmds}(e) and \ref{gtmds}(f), these GTMDs at fixed value of  $\textbf{p}_\perp=0.3$ $GeV$ with different values of $\bf{\Delta}_\perp$ are shown. In this model, $F_{1,2}$ vanishes. The distributions $F_{1,1}$, $F_{1,3}$ and $F_{1,4}$ corresponds to the Wigner distributions named $\rho_{UU}$, $\rho^i_{TU}$ and $\rho_{LU}$ respectively. In the context of longitudinal momentum fraction $x$, $F_{1,1}$ shows the maximum distribution at $\textbf{p}_{\perp}=0.1$ $GeV$ and the distribution peaks shift towards the lower values of $x$ with increasing the quark momentum in the transverse direction $\textbf{p}_\perp$ while in case of fixed $\textbf{p}_\perp$, the distribution peaks shift toward the higher $x$ with increasing the total momentum transferred to the proton $\bf \Delta_{\perp}$. From Eq. (\ref{f13}), $F_{1,3}$ shows the contribution of $F_{1,1}$. In case of $F_{1,4}$, Fig. {\ref{gtmds}}(c) shows the maximum distribution at $\textbf{p}_\perp=0.3$ $GeV$ while the peak moves towards the lower $x$ with increasing quark transverse momentum. 
%
%



\end{document}